\documentstyle[preprint,tighten,prc,aps,epsf]{revtex}

\begin{document}
\draft

\def\bra#1{{\langle#1\vert}}
\def\ket#1{{\vert#1\rangle}}
\def\coeff#1#2{{\scriptstyle{#1\over #2}}}
\def\undertext#1{{$\underline{\hbox{#1}}$}}
\def\hcal#1{{\hbox{\cal #1}}}
\def\sst#1{{\scriptscriptstyle #1}}
\def\eexp#1{{\hbox{e}^{#1}}}
\def\rbra#1{{\langle #1 \vert\!\vert}}
\def\rket#1{{\vert\!\vert #1\rangle}}
\def\lsim{{ <\atop\sim}}
\def\gsim{{ >\atop\sim}}
\def\nubar{{\bar\nu}}
\def\psibar{{\bar\psi}}
\def\Gmu{{G_\mu}}
\def\alr{{A_\sst{LR}}}
\def\wpv{{W^\sst{PV}}}
\def\evec{{\vec e}}
\def\notq{{\not\! q}}
\def\notk{{\not\! k}}
\def\notp{{\not\! p}}
\def\notpp{{\not\! p'}}
\def\notder{{\not\! \partial}}
\def\notcder{{\not\!\! D}}
\def\notA{{\not\!\! A}}
\def\notv{{\not\!\! v}}
\def\Jem{{J_\mu^{em}}}
\def\Jana{{J_{\mu 5}^{anapole}}}
\def\nue{{\nu_e}}
\def\mn{{m_\sst{N}}}
\def\mns{{m^2_\sst{N}}}
\def\me{{m_e}}
\def\mes{{m^2_e}}
\def\mq{{m_q}}
\def\mqs{{m_q^2}}
\def\mz{{M_\sst{Z}}}
\def\mzs{{M^2_\sst{Z}}}
\def\ubar{{\bar u}}
\def\dbar{{\bar d}}
\def\sbar{{\bar s}}
\def\qbar{{\bar q}}
\def\sstw{{\sin^2\theta_\sst{W}}}
\def\gv{{g_\sst{V}}}
\def\ga{{g_\sst{A}}}
\def\pv{{\vec p}}
\def\pvs{{{\vec p}^{\>2}}}
\def\ppv{{{\vec p}^{\>\prime}}}
\def\ppvs{{{\vec p}^{\>\prime\>2}}}
\def\qv{{\vec q}}
\def\qvs{{{\vec q}^{\>2}}}
\def\xv{{\vec x}}
\def\xpv{{{\vec x}^{\>\prime}}}
\def\yv{{\vec y}}
\def\tauv{{\vec\tau}}
\def\sigv{{\vec\sigma}}
\def\sst#1{{\scriptscriptstyle #1}}
\def\gpnn{{g_{\sst{NN}\pi}}}
\def\grnn{{g_{\sst{NN}\rho}}}
\def\gnnm{{g_\sst{NNM}}}
\def\hnnm{{h_\sst{NNM}}}

\def\xivz{{\xi_\sst{V}^{(0)}}}
\def\xivt{{\xi_\sst{V}^{(3)}}}
\def\xive{{\xi_\sst{V}^{(8)}}}
\def\xiaz{{\xi_\sst{A}^{(0)}}}
\def\xiat{{\xi_\sst{A}^{(3)}}}
\def\xiae{{\xi_\sst{A}^{(8)}}}
\def\xivtez{{\xi_\sst{V}^{T=0}}}
\def\xivteo{{\xi_\sst{V}^{T=1}}}
\def\xiatez{{\xi_\sst{A}^{T=0}}}
\def\xiateo{{\xi_\sst{A}^{T=1}}}
\def\xiva{{\xi_\sst{V,A}}}

\def\rvz{{R_\sst{V}^{(0)}}}
\def\rvt{{R_\sst{V}^{(3)}}}
\def\rve{{R_\sst{V}^{(8)}}}
\def\raz{{R_\sst{A}^{(0)}}}
\def\rat{{R_\sst{A}^{(3)}}}
\def\rae{{R_\sst{A}^{(8)}}}
\def\rvtez{{R_\sst{V}^{T=0}}}
\def\rvteo{{R_\sst{V}^{T=1}}}
\def\ratez{{R_\sst{A}^{T=0}}}
\def\rateo{{R_\sst{A}^{T=1}}}

\def\mro{{m_\rho}}
\def\mks{{m_\sst{K}^2}}
\def\mpi{{m_\pi}}
\def\mpis{{m_\pi^2}}
\def\mom{{m_\omega}}
\def\mphi{{m_\phi}}
\def\Qhat{{\hat Q}}

\def\FOS{{F_1^{(s)}}}
\def\FTS{{F_2^{(s)}}}
\def\GAS{{G_\sst{A}^{(s)}}}
\def\GES{{G_\sst{E}^{(s)}}}
\def\GMS{{G_\sst{M}^{(s)}}}
\def\GATEZ{{G_\sst{A}^{\sst{T}=0}}}
\def\GATEO{{G_\sst{A}^{\sst{T}=1}}}
\def\mdax{{M_\sst{A}}}
\def\mustr{{\mu_s}}
\def\rsstr{{r^2_s}}
\def\rhostr{{\rho_s}}
\def\GEG{{G_\sst{E}^\gamma}}
\def\GEZ{{G_\sst{E}^\sst{Z}}}
\def\GMG{{G_\sst{M}^\gamma}}
\def\GMZ{{G_\sst{M}^\sst{Z}}}
\def\GEn{{G_\sst{E}^n}}
\def\GEp{{G_\sst{E}^p}}
\def\GMn{{G_\sst{M}^n}}
\def\GMp{{G_\sst{M}^p}}
\def\GAp{{G_\sst{A}^p}}
\def\GAn{{G_\sst{A}^n}}
\def\GA{{G_\sst{A}}}
\def\GETEZ{{G_\sst{E}^{\sst{T}=0}}}
\def\GETEO{{G_\sst{E}^{\sst{T}=1}}}
\def\GMTEZ{{G_\sst{M}^{\sst{T}=0}}}
\def\GMTEO{{G_\sst{M}^{\sst{T}=1}}}
\def\lamd{{\lambda_\sst{D}^\sst{V}}}
\def\lamn{{\lambda_n}}
\def\lams{{\lambda_\sst{E}^{(s)}}}
\def\bvz{{\beta_\sst{V}^0}}
\def\bvo{{\beta_\sst{V}^1}}
\def\Gdip{{G_\sst{D}^\sst{V}}}
\def\GdipA{{G_\sst{D}^\sst{A}}}
\def\fks{{F_\sst{K}^{(s)}}}
\def\FIS{{F_i^{(s)}}}
\def\fpi{{F_\pi}}
\def\fk{{F_\sst{K}}}

\def\RAp{{R_\sst{A}^p}}
\def\RAn{{R_\sst{A}^n}}
\def\RVp{{R_\sst{V}^p}}
\def\RVn{{R_\sst{V}^n}}
\def\rva{{R_\sst{V,A}}}
\def\xbb{{x_B}}

\def\lamtv{{\Lambda_\sst{TVPC}}}
\def\lamtvs{{\Lambda_\sst{TVPC}^2}}
\def\lamtvc{{\Lambda_\sst{TVPC}^3}}
\def\lamtvi{{\Lambda_\sst{TVPC}^{-1}}}
\def\lampv{{\Lambda_\sst{PV}}}

\def\osffp{{{\cal O}_7^{ff'}}}
\def\ospg{{{\cal O}_7^{\gamma g}}}
\def\ospz{{{\cal O}_7^{\gamma Z}}}

\def\PR#1{{{\em   Phys. Rev.} {\bf #1} }}
\def\PRC#1{{{\em   Phys. Rev.} {\bf C#1} }}
\def\PRD#1{{{\em   Phys. Rev.} {\bf D#1} }}
\def\PRL#1{{{\em   Phys. Rev. Lett.} {\bf #1} }}
\def\NPA#1{{{\em   Nucl. Phys.} {\bf A#1} }}
\def\NPB#1{{{\em   Nucl. Phys.} {\bf B#1} }}
\def\AoP#1{{{\em   Ann. of Phys.} {\bf #1} }}
\def\PRp#1{{{\em   Phys. Reports} {\bf #1} }}
\def\PLB#1{{{\em   Phys. Lett.} {\bf B#1} }}
\def\ZPA#1{{{\em   Z. f\"ur Phys.} {\bf A#1} }}
\def\ZPC#1{{{\em   Z. f\"ur Phys.} {\bf C#1} }}
\def\etal{{{\em   et al.}}}

\def\delalr{{{delta\alr\over\alr}}}
\def\pbar{{\bar{p}}}
\def\lamchi{{\Lambda_\chi}}
\def\gaf{{g_\sst{A}^f}}
\def\gvu{{g_\sst{V}^u}}
\def\gvd{{g_\sst{V}^d}}
\def\gve{{g_\sst{V}^e}}

\def\vuds{{|V_{ud}|^2}}
\def\vud{{V_{ud}}}
\def\vudssm{{|V_{ud}|^2_\sst{SM}}}
\def\vudsex{{|V_{ud}|^2_\sst{EX}}}
\def\qpv{{Q_\sst{W}}}
\def\qpvsm{{Q_\sst{W}^\sst{SM}}}
\def\qpvex{{Q^\sst{EX}_\sst{W}}}
\def\delqpv{{\Delta\qpv}}
\def\qwp{{Q_\sst{W}^p}}
\def\qwn{{Q_\sst{W}^n}}
\def\qwz{{Q_\sst{W}^0}}
\def\delqwp{{\Delta\qwp}}
\def\delqwn{{\Delta\qwn}}
\def\apv{{A_\sst{PV}}}
\def\mws{{M_\sst{W}^2}}

\def\deltar{{\Delta r}}
\def\deltarmu{{\Delta r_\mu}}
\def\deltamu{{\Delta_\mu}}

\title{Nuclear $\beta$-decay, Atomic Parity Violation,\\
and New Physics}

\author{
M.J. Ramsey-Musolf\\
\vspace{0.3cm}
}
\address{
Department of Physics, University of Connecticut\\
Storrs, CT 06269 USA\\
and \\
Theory Group, Thomas Jefferson National Accelerator Facility\\
Newport News, VA 23606 USA}


\maketitle

\begin{abstract}
Determinations of $\vuds$ with super-allowed Fermi $\beta$-decay in nuclei
and of the weak charge of the
cesium  in atomic parity-violation deviate from the Standard Model
predicitions by
$2\sigma$ or more. In both cases, the Standard Model over-predicts the
magnitudes of the relevant
observables. I discuss the implications of these results for R-parity
violating (RPV) extensions of the
minimal supersymmetric  Standard Model. I also explore the possible
consequences for RPV supersymmetry
of prospective future low-energy electroweak measurements.
\end{abstract}


\narrowtext

The search for physics beyond the Standard Model (SM) is one of the primary
objectives of present and
future high-energy collider experiments. At the same time, there exist a
variety of low- and
medium-energy  atomic and nuclear studies making important contributions in
the search for new physics.
For example, measurements of superallowed nuclear Fermi $\beta$-decay
provide the most precise determination of the $ud$ element of the
Cabibbo-Kobayashi-Maskawa (CKM) matrix,
$\vud$. When combined with determinations of $|V_{us}|$ and $|V_{ub}|$ from
$K$ and $B$ meson
decays \cite{PDG}, nuclear Fermi $\beta$-decay provides a stringent test of
CKM matrix unitarity. In the neutral current sector, the Boulder group has
obtained a precise
determination of the weak charge of the cesium atom, $\qpv$, using atomic
parity-violation (APV).
In both cases, the results deviate from the SM predictions by $2\sigma$ or
more. Denoting
$\vudssm$ the value implied by CKM unitarity and $\qpvsm$ the weak charge
computed in the SM, one has
\cite{Tow96,Hag96,Woo97,Ben99}:
\begin{eqnarray}\label{eq:exresults1}
(\vudsex - \vudssm)/\vudssm & = & -0.0029 \pm 0.0014 \\
\label{eq:exresults2}
(\qpvex-\qpvsm)/\qpvsm & = & -0.016 \pm 0.006 \ \ \ ,
\end{eqnarray}
where \lq\lq $EX$" denotes the experimental value for the corresponding
observable
and  where the experimental and systematic errors (including theoretical)
have been combined in
quadrature. The APV results correspond to a single experiment, whereas the
$\beta$-decay results have
been obtained by averaging over nine different decays. Interestingly, the
relative deviations from the
SM in both cases are negative.

Assuming the deviations in Eqs. (\ref{eq:exresults1}-\ref{eq:exresults2})
cannot be explained by
conventional hadronic, nuclear, or atomic effects, they may hint at the
presence of new physics. In
this respect, the cesium APV result has sparked considerable recent
attention. Among the more interesting
possbilities is that an additional neutral weak gauge boson is the culprit
behind the observed
deviation. The sign of the observed deviation has a natural explanation in
the context of
$E_6$ theories \cite{Lon86,MRM99,Cas00,Erl00}. The presence of an
additional U(1) symmetry alone,
however, would not help account for the longer-standing $\beta$-decay result.

In what follows, I investigate whether new physics scenarios exist which
might account for both
the common sign of the results in Eqs.
(\ref{eq:exresults1}-\ref{eq:exresults2}) as well as the
observed magnitudes. After making some general observations about the
impact of new interactions on
these observables, I illustrate using extensions of the minimal
supersymmetric SM having
R-parity violating (RPV) interactions. I show that low-energy electroweak
data place severe
constraints on this scenario. Nevertheless, at the 2$\sigma$ level, there
exists a small but
non-vanishing region in the parameter space of RPV couplings and sfermiom
masses which may account for
the $\beta$-decay and APV results. I also show that, within this framework,
consistency of the
low-energy results with  rare decay limits does not appear to require
significant mass hierarchies in
the sfermion spectrum. Finally, if RPV supersymmetry is responsible for the
results
in Eqs. (\ref{eq:exresults1}-\ref{eq:exresults2}), observable consequences
may also follow for other
prospective low-energy precision measurements. I discuss three such cases:
(i) a measurement of
the PV M\"oller scattering asymmetry, (ii) a determination of the weak
charge of the proton using
parity-violating electron scattering (PVES), and (ii) a measurement of
ratios of APV observables for
different atoms along an isotope chain. The sensitivity of all three
measurements to new RPV interactions
differs substantially from that of $\beta$-decay and APV. I discuss the
conditions under which these
new measurements may impose further constraints on the RPV parameter space.

In general, the presence of new physics may modify low-energy semileptonic
electroweak observables in
two ways: (i) directly, via a new semileptonic interaction or modification
of the SM semileptonic
interaction, and (ii) indirectly, through a modification of the relative
normalizations of leptonic and
semileptonic amplitudes. Indirect effects may arise because semileptonic SM
amplitudes are expressed in
terms of $G_\mu$, the Fermi constant measured in $\mu$-decay. In the SM, it
is related to the semiweak
couplings as
\begin{equation}
\label{eq:GFSM}
{G_\mu^\sst{SM}\over\sqrt{2}} = {g^2\over 8\mws} + {\hbox{rad. corr.}} \equiv
{g^2\over 8\mws}[1+\deltarmu]  \ \ \ ,
\end{equation}
where \lq\lq rad. corr." and $\deltarmu$ denote the appropriate radiative
corrections to the tree-elvel
$\mu$-decay amplitude. The presence of new leptonic physics modifies the
relation (\ref{eq:GFSM}) as
\begin{equation}
\label{eq:GFnew}
{G_\mu\over\sqrt{2}} = {g^2\over 8\mws}\left(1+\deltarmu +\deltamu\right) =
{G_\mu^\sst{SM}\over\sqrt{2}} (1+\deltamu)\ \ \ ,
\end{equation}
where $\deltamu$ denotes the new physics correction to the tree-level SM
$\mu$-decay amplitude.
When the SM is used to compute $\beta$-decay or APV amplitudes, one
requires $g^2/\mws$ as input. Since
$G_\mu$ is one of the three most precisely measured electroweak input
parameters, it is standard to
rewrite $g^2/\mws$ in terms of $G_\mu$ using Eq. (\ref{eq:GFSM}). Thus, the
presence of
$\deltamu$ would modify the normalization of the $\beta$-decay and APV
amplitudes via Eq.
(\ref{eq:GFnew}).

In the case of PV neutral current amplitudes, an additional
$\deltamu$-dependence arises from the
determination of the weak mixing angle. At tree-level in the SM, the weak
charge is given by
\begin{equation}
\label{eq:qwtree}
\qwz = Z(1-4x) -N \ \ \ ,
\end{equation}
where $x\equiv\sstw$ is computed in terms of $\alpha$, $G_\mu$, and $\mz$
from the relation
\begin{equation}
\label{eq:sstwrel}
x(1-x) = {\pi\alpha\over\sqrt{2} G_\mu\mzs(1-\deltar-\deltamu)} \ \ \ ,
\end{equation}
and where the precise values of $x$ and the radiative corrections $\deltar$
depend on the choice of
renormalization scheme. The $\deltamu$-dependence of $\sstw$ in Eq.
(\ref{eq:sstwrel}) translates
into a corresponding dependence of $\deltamu$ in $\qpv$.

In order to delineate the effects of new leptonic and semileptonic physics
in the semileptonic
observables of interest here, it is useful to define effective Fermi
constants for the latter:
\begin{eqnarray}
\label{eq:GFbeta}
G_F^\beta & = & G_\mu
|V_{ud}|\left(1-\deltarmu+\deltar_\beta-\deltamu+\Delta_\beta\right) \\
\label{eq:GFpv}
G_F^\sst{PV} & = & G_\mu
\qwz(\deltamu)\left(1-\deltarmu+\deltar_\sst{PV}-\deltamu+\Delta_\sst{PV}
\right) \ \ \ ,
\end{eqnarray}
where $\deltar_\beta$ and $\deltar_\sst{PV}$ denote the appropriate SM
radiative corrections to
the charged current $\beta$-decay and neutral current PV amplitudes,
respectively, and where
$\Delta_\beta$ and $\Delta_\sst{PV}$ denote the corresponding semileptonic
new physics corrections.
The $\deltamu$-dependence of $\qwz$ arises for the reasons discussed
above\footnote{It is conventional to
define the SM weak charge as $Q_\sst{W}^\sst{SM}=
\qwz(1-\deltarmu+\deltar_\sst{PV})$.}.
The experimental results imply that
\begin{eqnarray}
\label{eq:GFbetaex}
G_F^{\beta, \sst{EX}}/G_F^{\beta, \sst{SM}} &< & 1 \\
\label{eq:GFpvex}
G_F^{\sst{PV}, \sst{EX}}/G_F^{\sst{PV}, \sst{SM}}& < & 1\ \ \ ,
\end{eqnarray}
where the SM values are computed using $\deltamu = \Delta_\beta =
\Delta_\sst{PV} = 0$. The
conventional interpretation of the reduction in effective Fermi constants
is given in
Eqs. (\ref{eq:exresults1}-\ref{eq:exresults2}).

At first
glance, it appears that a positive value for $\deltamu$ would reduce the
effective Fermi constants
from their SM values and explain the sign of the observed deviations
without requiring the
interpretation of Eqs. (\ref{eq:exresults1}-\ref{eq:exresults2}).
In the case of APV, however,
the $\deltamu$-dependence of $\qwz$ cancels against the $\deltamu$-induced
modification of the overall
normalization, yielding a negligible net effect from $\deltamu$ on
$G_F^\sst{PV}$.\footnote{This
cancellation was first noted in Ref. \cite{Mar90} in the context of oblique
corrections to electroweak
observables.} To see this cancellation explicitly, one may expand
$\qwz(\deltamu)$ to first order in
$\deltamu$ using Eq. (\ref{eq:sstwrel}), yielding
\begin{equation}
G_F^\sst{PV} \approx G_\mu \qwz \left(1-\deltarmu+\deltar_\sst{PV} +
\xi\deltamu +\Delta_\sst{PV}
\right) \ \ \ ,
\end{equation}
where
\begin{eqnarray}
\label{eq:xi}
\xi &=&-1-(4Z/\qwz)\lambda_x \\
\label{eq:lambdax}
\lambda_x &\approx& {x(1-x)\over 1-2x}{1\over 1-\deltar} \ \ \ .
\end{eqnarray}
For cesium, $\xi\approx 0.05$ when the weak mixing angle is defined in the
$\overline{MS}$ scheme. Thus, while a non-zero value for $\deltamu$ might
account for the reduction in
$G_F^\beta$ from its SM value, it is an unlikely source of the 1.6\%
reduction in $G_F^\sst{PV}$.
Instead, one must look to new semileptonic neutral current interactions to
generate the observed
APV effect.

Extensions of the minimal supersymmetric standard model (MSSM) containing
RPV interactions can
generate tree-level contributions to $\deltamu$, $\deltar_\beta$, and
$\deltar_\sst{PV}$.
The MSSM is a popular candidate for SM extensions. Although no direct
evidence for supersymmetry
(SUSY) has yet been obtained, there exist  compelling theoretical arguments
as to why it should be
correct (for a review, see Ref. \cite{Mar98}). The MSSM can be extended to
include terms in the
superpotential which do not conserve the quantum number
$P_R = (-1)^{3(B-L)+2S}$, where $B$ and $L$ denote baryon and lepton
number, respectively, and $S$
is the spin of a given particle. Such RPV interactions result in the
Lagrangians
\cite{Bar89}
\begin{eqnarray}
\label{eq:lsusy}
{\cal L}_\sst{RPV}& = & \lambda_{ijk}[{\tilde\nu}^i_L {\bar e}^k_R e^j_L +
{\tilde e}^j_L{\bar e}^k_R
    \nu^i_L+ ({\tilde e}^k_R)^{\ast} ({\bar\nu}^i_L)^c e^j_L \\ \nonumber
  && \ \ \ -(i\leftrightarrow j)] + {\hbox{h.c.}} \\ \nonumber
&& + \lambda_{ijk}'[{\tilde\nu}_L^i {\bar d}^k_R d_L^j +{\tilde d}^j_L
{\bar d}^k_R\nu^i_L
   + ({\tilde d}^k_R)^{\ast} ({\bar\nu}^i_L)^c d^j_L \\ \nonumber
&& \ \ \ - {\tilde e}_L^i {\bar d}^k_R u_L^j -{\tilde u}^j_L {\bar d}^k_R e^i_L
   - ({\tilde d}^k_R)^{\ast} ({\bar e}^i)^c_L u^j_L] + {\hbox{h.c.}}\ \ \ ,
\end{eqnarray}
where the $i,j,k$ indices denote generation and where the $\tilde f$
denotes the supersymmetric partner
of the corresponding fermion $f$. Both the $\lambda$ and $\lambda'$ terms
in Eq. (\ref{eq:lsusy}) violate
lepton number conservation.

At low-energies, the exchange of a sfermion between SM fermions yields
four-fermion effective
interactions. Upon Fierz reordering, these interactions take on the
structure of the corresponding
effective current-current interactions in the SM. Consequently, one expects
${\cal L}_\sst{RPV}$ to
induce corrections to low-energy electroweak observables. In the present
context, one may express
these corrections in terms of the quantities $\Delta_{12k}({\tilde
e_R^k})$, $\Delta_{11k}'({\tilde
d_R^k})$, $\Delta_{1j1}'({\tilde q_L^j})$, where
\begin{equation}
\label{eq:deltadef}
\Delta_{12k}({\tilde e_R^k}) = {|\lambda_{12k}|^2\over
4\sqrt{2}G_\mu^\sst{SM}M^2_{\tilde e_R^k}} \ \ \ .
\end{equation}
with ${\tilde e_R^k}$ being the exchanged slepton, and where
$\Delta_{11k}'({\tilde d_R^k}) $ and $\Delta_{1j1}'({\tilde q_L^j})$ are
defined as in Eq.
(\ref{eq:deltadef}) but with $\lambda_{12k}\to\lambda_{11k}'$, $
M^2_{\tilde e^k_R}\to M^2_{\tilde
d^k_R}$ and $\lambda_{12k}\to\lambda_{1j1}'$, $ M^2_{\tilde e^k_R}\to
M^2_{\tilde q^j_L}$, respectively.
In terms of these quantities, which are non-negative, one has
\begin{eqnarray}
\label{eq:susycontrib1}
\Delta_\beta-\deltamu &\approx & \Delta_{11k}'({\tilde
d_R^k})-\Delta_{12k}({\tilde e_R^k}) \\
\label{eq:susycontrib2}
\Delta_\sst{PV}+\xi\deltamu & \approx & 0.05\Delta_{12k}({\tilde e_R^k}) -
2\left({2Z + N\over
N}\right)\Delta_{11k}'({\tilde d_R^k}) \\
  && \ \  + 2\left({2N+Z\over N}\right)\Delta_{1j1}'({\tilde q_L^j}) \ \ \
. \nonumber
\end{eqnarray}
In arriving at the expression  in Eqs. (\ref{eq:susycontrib2}) I have
omitted small
contributions to the tree-level amplitude involving $1-4\sstw$. Note that
$\Delta_{11k}'$ and $\Delta_{12k}$ cancel  against each other in the
$\beta$-decay amplitude. In contrast, the impact of $\Delta_{12k}$ on the
PV amplitude is
suppressed while the effects of the $\lambda'$ terms are enhanced by the
factors
$2(2Z+N)/N\sim 2(2N+Z)/N\sim 5$.

Typically, limits on the RPV interactions of Eqs. (\ref{eq:lsusy}) are
obtained assuming all but one of
the $\lambda_{ijk}$ and $\lambda_{ijk}'$ vanish. In the present case,
however, a common explanation for
the $\beta$-decay and APV results does not obtain if only one of the terms
in Eq. (\ref{eq:lsusy}) is
non-vanishing. For example, taking $\Delta_{12k}>0$ but
$\Delta_{11k}'=0=\Delta_{1j1}'$ could not
account for the common sign of both the $\beta$-decay and APV deviations.
Similarly, taking $\Delta_{12k}=0$ but either
$\Delta_{11k}'\not= 0$ or $\Delta_{1j1}' \not= 0$ would not generate the
observed phases\footnote{A
recent analysis of APV and other semi-leptonic data in terms of
leptoquark interactions has been reported in Ref. \cite{Bar00}. In that
analysis, no new purely leptonic
interactions were included. These authors find -- as noted here -- that the
APV and charged current
decay results are not consistent with $\Delta'_{11k} >0$ in the absence of
new leptonic physics.}. A
potentially successful scenario may arise when the both a leptonic and a
semi-leptonic RPV interaction
occur.

To illustrate,  consider the case in which $\Delta_{12k}>\Delta_{11k}'
>\Delta_{1j1}'= 0$. In Fig. 1 I show the values of these corrections needed
>to account for
the low-energy results at the 2$\sigma$ level. By themselves, these results
allow $\Delta_{12k}$
and $\Delta_{11k}'$ to differ from zero over considerable ranges. A further
restriction on the
allowed region is obtained by studying the results of $\pi_{\ell 2}$
decays. The ratio
\begin{equation}
\label{eq:piratio1}
R_{e/\mu} = {\Gamma(\pi^+\to e^+\nu_e + \pi^+\to
e^+\nu_e\gamma)\over\Gamma(\pi^+\to \mu^+\nu_\mu +
\pi^+\to \mu^+\nu_\mu\gamma)}
\end{equation}
has been measured precisely at PSI \cite{Cza93} and TRIUMF \cite{Bri92}.
Comparing the Particle
Data Group average \cite{PDG} with the SM value as calculated in Ref.
\cite{Mar93} one has
\begin{equation}
\label{eq:piratio2}
{R_{e/\mu}^\sst{EX}\over R_{e/\mu}^\sst{SM}} = 0.9958\pm 0.0033 \pm 0.0004
\end{equation}
where the first error is experimental and the second is theoretical. In
terms of RPV interactions, one
has
\begin{equation}
\label{eq:piratio3}
{R_{e/\mu}\over R_{e/\mu}^\sst{SM}} = 1 + 2\left[\Delta_{11k}'({\tilde
d_R^k})-\Delta_{21k}'({\tilde
d_R^k})\right] \ \ \ .
\end{equation}
Note that the leptonic correction $\Delta_{12k}$ to the overall
normalization cancels from the ratio of
these charged current decays, leaving only the new semileptonic
contributions. Assuming
$\Delta_{11k}'({\tilde d_R^k})>\Delta_{21k}'({\tilde d_R^k})= 0$ one
obtains strong upper bounds on
$\Delta_{11k}'({\tilde d_R^k})$ from the results in Eq.
(\ref{eq:piratio3}). The corresponding $2\sigma$
bounds are also shown in Fig. 1.

In principle, an additional restriction on the allowed region arises from the
the self-consistency of electroweak parameters.  For example,
one may relate $G_F^\sst{SM}$ to other parameters in the SM \cite{WJM75,Sir80}
\begin{equation}
\label{eq:ewkrelation}
G_F^\sst{SM} =
{\pi\alpha\over\sqrt{2}\mws\sstw(\mz)_{\overline{MS}}(1-\Delta
r(\mz)_{\overline{MS}})}
\ \ \ ,
\end{equation}
where $\Delta r(\mz)_{\overline{MS}}$ denotes a radiative correction to
this relation in the
$\overline{MS}$-scheme. From a comparison of $G_\mu$ with the value of the
Fermi constant computed
according to Eq. (\ref{eq:ewkrelation}), one obtains the $2\sigma$
limits\footnote{Note that
$\Delta_{12k}\geq 0$ according to Eq. (\ref{eq:deltadef}).}
\footnote{For a similar analysis in terms of the oblique parameters, see Ref.
\cite{WJM99}}
\begin{equation}
\label{eq:deltaconstraint}
-0.0035 < \Delta_{12k} < 0.0040 \ \ \ .
\end{equation}
This constraint is also shown in Fig. 1 (similar constraints can be
obtained in other
renormalization schemes.) At this level, the bounds from Eq.
(\ref{eq:deltaconstraint}) do not
significantly impact the allowed region. The approximate centroid of the
allowed is given by
($\Delta_{12k}=0.0025$, $\Delta_{11k}'=0.0010$). This point corresponds to
a -0.15\% shift in $G_F^\beta$
and a -0.5\% change in $G_F^\sst{PV}$ from the SM values.

In general, experimental limits on flavor-changing neutral currents and
other rare processes
impose stringent limits on products of the $\lambda_{ijk}$ and
$\lambda_{ijk}'$ couplings when two or
more are simultaneously non-vanishing. The case considered above is no
exception. However, when the
purely leptonic correction $\Delta_{12k}(\tilde e_R^k)$ involves the
exchange of a $\tau$ slepton
($k=3$), the limits from rare processes do not appear to rule out the
simultaneous occurrence of a
leptonic and semi-leptonic RPV interaction. For example, if the
$\lambda_{123}$ and $\lambda_{11k}'$
($k=2$ or 3 but not both) interactions are both non-zero, then the decays
$B^0\to\tau^{\pm}\mu^{\pm}$
(k=3) or
$\tau\to\mu K^0$ (k=2) can occur via the exchange of a ${\tilde\nu}_{e_L}$.
The corresponding branching
ratios are
\cite{PDG} (90\% C.L.)
\begin{eqnarray}
B(\tau\to\mu K^0) & < & 1\times 10^{-3} \\
B(B^0\to \tau^{\pm}\mu{\pm}) &  < & 8.3 \times 10^{-4} \ \ \ .
\end{eqnarray}
These results imply that
\begin{eqnarray}
\label{eq:limits1}
\sqrt{|\lambda_{112}'\lambda_{123}|} & < 0.11\  (M_{\tilde\nu_e^L}/100\
{\hbox{GeV}}) \\
\label{eq:limits1b}
\sqrt{|\lambda_{113}'\lambda_{123}|} & < 0.036\  (M_{\tilde\nu_e^L}/100\
{\hbox{GeV}}) \ \ \ .
\end{eqnarray}
By comparison, taking $\Delta_{12k}({\tilde e_R^k})=0.0025$ and
$\Delta_{11k}'({\tilde d_R^k})=
0.0010$ as above would require
\begin{eqnarray}
\label{eq:limits2}
|\lambda_{12k}| & =&  0.041\  (M_{\tilde e_R^k}/100\ {\hbox{GeV}})\\ \nonumber
|\lambda_{11k}'| & =&  0.026\  (M_{\tilde d_R^k}/100\ {\hbox{GeV}})\ \ \ .
\end{eqnarray}
If $M_{\tilde\nu_e^L}\sim M_{\tilde e_R^k}\sim M_{\tilde d_R^k}$, then Eqs.
(\ref{eq:limits2}) imply
\begin{equation}
\label{eq:limits3}
\sqrt{|\lambda_{11k}'\lambda_{12k}|} \sim 0.033\  (M_{\tilde f}/100\
{\hbox{GeV}}) \ \ \ ,
\end{equation}
where $M_{\tilde f}$ is a common sfermion mass scale. Comparing Eqs.
(\ref{eq:limits3})
and (\ref{eq:limits1},\ref{eq:limits1b}), one sees that the $\beta$-decay
and APV results and
rare decay limits can be accomodated in the RPV MSSM without requiring mass
heirarchies in the soft
SUSY-breaking sector.

The viability of RPV supersymmetry in the present context would be further
constrained by improved
limits on rare $B$ and $\tau$ decays. In the light flavor sector, it may
be tested by future low-energy electroweak measurements. New measurements
of pion, neutron, and Fermi
nuclear $\beta$-decay will further test the deviation of $G_F^\beta$ from
the SM value. A new
determination of the $^{10}{\hbox{C}}(0^+, {\hbox{g.s.}})\to ^{10}
{\hbox{B}}(0^+, 1.74\ {\hbox{MeV}})$ branching
ratio
\cite{Fuj99} yields a value for $G_F^\beta$ consistent with the SM value,
though the errors are
considerably larger than those corresponding to Eq. (\ref{eq:exresults1}).
A 0.7\% determination of
the neutron $\beta$-decay asymmetry parameter $A$ has been obtained at ILL
\cite{Rei99}. When combined
with the world average for the neutron lifetime, the new value for $A$
implies an even smaller value
for $G_F^\beta$ than obtained from the average of superallowed decays, with
a similar uncertainty.
A future, precise determination of $A$ is underway at Los Alamos.

Among neutral current studies, a PV M\"oller scattering experiment is
planned for SLAC \cite{Kum97}. The
M\"oller asymmetry is sensitive to the leptonic correction $\Delta_{12k}$.
At tree level, one has
\begin{equation}
\label{eq:moller}
\delta_e =  \alr(ee)/A_\sst{LR}^\sst{SM}(ee) \approx -\left[1+\left({4\over
1-4\sstw}\right)\lambda_x\right]
\Delta_{12k}(\tilde e^k_R)\ \ \ .
\end{equation}
Including the ${\cal O}(\alpha)$ electroweak corrections in
$A_\sst{LR}^\sst{SM}$ \cite{Cza96} leads to
$\delta_e\approx - 31\Delta_{12k}(\tilde e^k_R)$. The expected precision
for this experiment is
$\pm 7\%$. Thus, a result implying $\delta_e^\sst{EX}\gsim 0.11$ would
begin to impact the 2$\sigma$
constraints in Fig. 1.

In the semi-leptonic sector, additional experiments are planned in APV.
These measurements will consider
ratios of PV observables along an isotope chain in order to reduce the
effect of atomic theory
uncertainties. For example, if $A_\sst{PV}(N)$ denotes an APV observable
for an isotope with $N$
neutrons, one may consider
\begin{equation}
{\cal R}=
{\apv(N')-\apv(N)\over\apv(N')+\apv(N)}\approx{\qpv(N')-\qpv(N)\over\qpv
(N')+\qpv(N)} \ \ \ .
\end{equation}
Letting ${\cal R}={\cal R}^\sst{SM}(1+\delta_{\cal R})$, where ${\cal
R}^\sst{SM}$
denotes the value in the SM, one has \cite{MRM99,MRM00}
\begin{eqnarray}
\label{eq:rshift}
\delta_{\cal R}&\approx & 2\left({2Z\over N'+
N}\right)[-2\lambda_x\Delta_{12k}(\tilde e^k_R)+
2\Delta_{11k}'(\tilde d_R^k)-\Delta_{1j1}'(\tilde
q_L^j)] \\
&& \ \ \  -
  \left({N'\over\Delta N}\right) (Z\alpha)^2(3/7)\delta(\Delta X_N) \ \ \ .
\nonumber
\end{eqnarray}
Here, I have followed Refs. \cite{For90,Pol92} and approximated the nucleus
as a sphere of constant
neutron and proton densities out to radii $R_N$ and $R_P$, respectively.
The parameter $\Delta X_N =
(R_{N'}-R_N)/R_P$ and $\delta\Delta X_N$ denotes the uncertainty in this
quantity. Note that unlike the
correction to the PV amplitude for a single isotope, the dependence of
$\delta_{\cal R}$ on the purely
leptonic new physics is not negligible. Given the allowed region in Fig. 1,
the first term in Eq. (\ref{eq:rshift}) could range between -0.0006 and
0.0019.
Although one anticipates an experimental uncertainty in $\delta_{\cal R}$
of $\sim 0.001 - 0.003$, the
uncertainty in the nuclear structure term is likely to be larger
\cite{MRM99}. The sensitivity of
isotope ratio measurements to possible RPV effects is thus complicated by
nuclear structure
uncertainty.

Alternatively, one may access the RPV corrections with a PV electron
scattering (PVES) measurement of the
proton's weak charge. The relative shift induced in this case is
\begin{equation}
\label{eq:proton}
\delta_P = \Delta\qwp/\qwp\approx \left({2\over
1-4\sstw}\right)[-2\lambda_x\Delta_{12k}(\tilde e_R^k) +
2\Delta_{11k}'(\tilde d_R^k)-\Delta_{1j1}'(\tilde q_L^j)] \ \ \ ,
\end{equation}
where a small contribution to the coefficient of $\Delta_{12k}$
proportional to $(1-4\sstw)$
has been omitted for simplicity of illustration.
Note that -- apart from the latter -- the dependence of $\qwp$ on new RPV
physics is the same as that
of ${\cal R}$, to first order in the new interactions. This feature is
general and applies to situations
other than the RPV SUSY scenario discussed here \cite{MRM99}. From Eq.
(\ref{eq:proton}), one would
expect $-0.03 \leq \Delta\qwp/\qwp\leq 0.4$ for the allowed region in Fig.
1. Alternatively, a
3\% determination of $\qwp$ would begin to tighten the 2$\sigma$ allowed
region if
$\delta_P^\sst{EX}\lsim -0.02$. Recently, a letter of intent to measure
$\qwp$ at the 3-5\% level with
PVES at the Jefferson Lab has appeared \cite{LOI}.  In contrast to the
situation with the isotope
ratios, the interpretation of a 3\% PVES determination $\qwp$ does not
appear to be limited by strong
interaction uncertainties. Such a measurement could place new and
interesting constraints on the
possibility of low-energy RPV effects.

\acknowledgements

It is a pleasure to thank R. Lebed for useful discussions, W. Marciano for
comments on an earlier
version of this manuscript, and for S.J. Puglia for assistance in preparing
the figure. This work was
supported in part under U.S. Department of Energy contract
\#DE-AC05-84ER40150 and a National Science Foundation Young Investigator
Award.

\vfill
\eject

\begin{figure}
\caption{\label{Fig1} The $2\sigma$ constraints on RPV corrections
$\Delta_{12k}(\tilde e_R^k)$ and
$\Delta_{11k}'(\tilde d_R^k)$ from precision electroweak data. Dark solid
lines give constraints from
superallowed nuclear $\beta$-decay. Dashed lines indicate APV constraints,
while light vertical solid
line corresponds to bounds of Eq. (\ref{eq:deltaconstraint}). Dot-dashed
line gives upper bound from
$\pi_{\ell 2}$ decays. The allowed region is indicated by
shading.}
\end{figure}

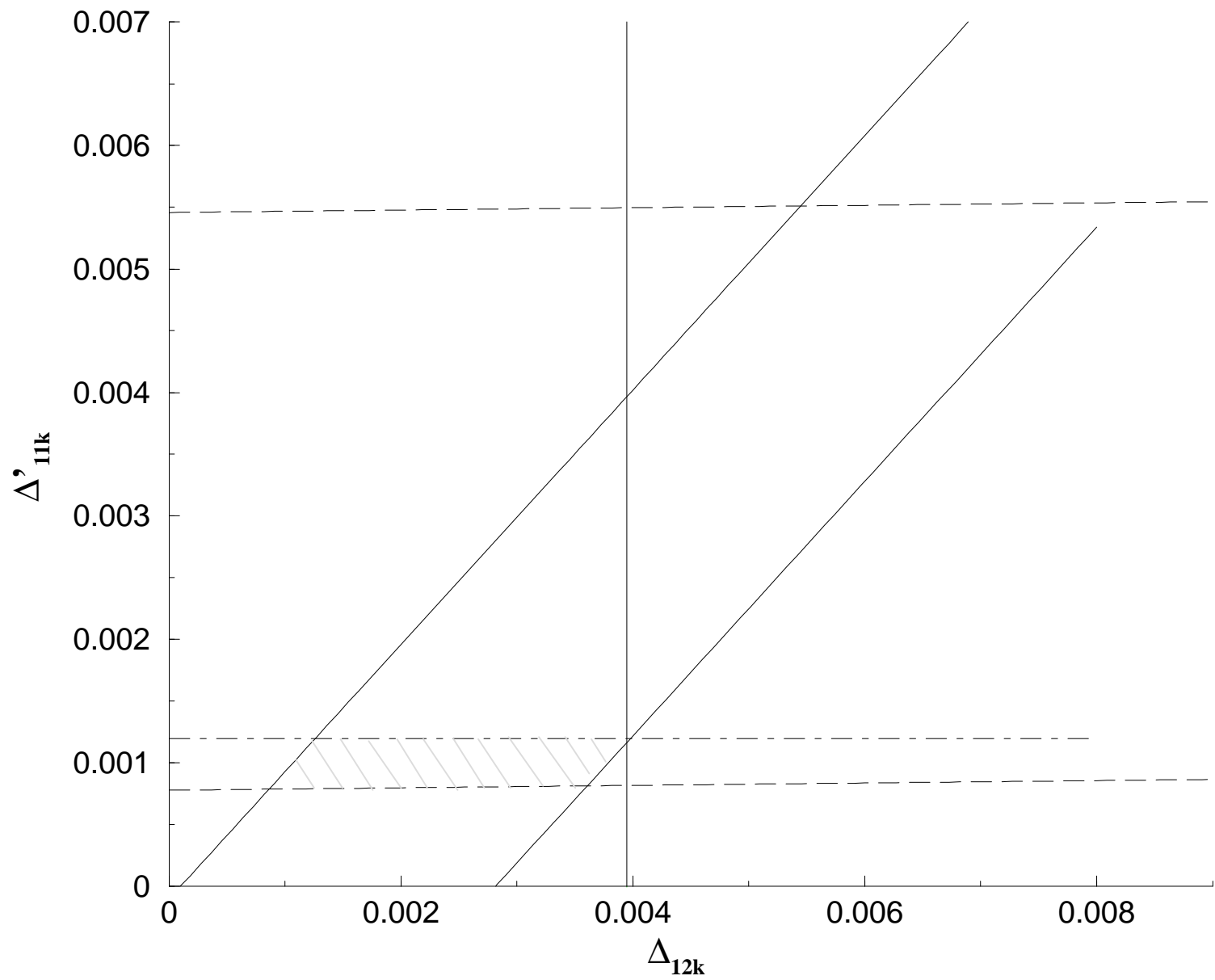
\end{document}